\documentclass[conference]{IEEEtran}
\IEEEoverridecommandlockouts

\usepackage{amsmath,amssymb}
\usepackage[hidelinks]{hyperref}
\usepackage[capitalize]{cleveref}
\usepackage{orcidlink}
\usepackage{booktabs}
\usepackage{subcaption}
\usepackage[dvipsnames, x11names, svgnames]{xcolor}
\usepackage{tikz}
\usepackage{pgfplots}
\usepackage{siunitx}
\usepackage{cite}
\usepackage{ifthen}  

\crefname{figure}{Fig.}{Figs.}
\crefname{table}{Tab.}{Tabs.}

\usetikzlibrary{positioning, calc, shapes.geometric, shapes.symbols, fit, external, arrows.meta}
\pgfplotsset{compat = newest}

\usepackage{fancyhdr}
\newcommand{\preprintheader}{{\footnotesize\scshape Preprint, \today}}
\fancypagestyle{plain}{%
  \fancyhf{}%
  \fancyhead[R]{\preprintheader}%
}

\begin{document}

\title{End-to-End Modelling of Earthquake-Induced Polarisation Perturbations in Submarine Optical Fibres}

\author{Rick M. Butler\textsuperscript{\textdagger,\textdaggerdbl,*},
Jos\'e N\'u\~nez-Kasaneva\textsuperscript{\textdagger},
Gabriele Liga\textsuperscript{\textdagger},
Magnus Karlsson\textsuperscript{\textdaggerdbl},\\
Alex Alvarado\textsuperscript{\textdagger},
Christian H\"ager\textsuperscript{\textdaggerdbl} \\
\textit{\textsuperscript{\textdagger}Department of Electrical Engineering, Chalmers University of Technology, Gothenburg, Sweden} \\
\textit{\textsuperscript{\textdaggerdbl}Department of Electrical Engineering, Eindhoven University of Technology, Eindhoven, The Netherlands} \\
\textsuperscript{*}rick.butler@chalmers.se
}

\maketitle
\thispagestyle{fancy}  

\begin{abstract}
We present an end-to-end model of earthquake-perturbed signal propagation in submarine fibres. Ground displacements are converted to fibre strain via ocean pressure, and injected into a split-step model. Our model shows good qualitative alignment with measurements from the Curie cable during the 2020 Oaxaca earthquake.
\end{abstract}

\section{Introduction}
Earthquake early warning systems \cite{Cremen:Jun20:earthquake_warning_advances} traditionally rely on seismometer networks \cite{Ringler:Jul22:achievements_seismographic_networks}.
However, offshore seismometer presence is sparse, whereas optical communication fibres are widespread \cite{submarine_cable_map}.
Therefore, it has been proposed to use existing submarine telecom fibres for earthquake sensing \cite{Lindsey:Nov17:fiber_observations_earthquake,Marra:Jun18:interferometry_earthquake_detection,Jousset:Jul18:strain_fibre_seismological}. Indeed, earthquakes induce spatiotemporal perturbations of the fibre geometry \cite{Mecozzi:Jun21:polarization_sensing_submarine}.
Fibre strain affects the birefringence, i.e., the polarisation-dependence of the speed of light in the fibre.
Hence, perturbations can be sensed from variations in the received state of polarisation (SOP).

Sensing algorithm design requires earthquake-perturbed fibre output data. 
Capturing the resulting SOP variations in measurements is challenging \cite{Geller:Mar97:earthquakes_cannot_predicted}, and only a few experimental datasets are publicly available, e.g., \cite{Zhan:Feb21:polarization_wave_sensing,Carver:Jul24:polarization_sensing_seismic}. 
Alternatively, given an earthquake event, fibre link, and input signal, a numerical propagation model could provide controlled and reproducible fibre output SOP traces.
Such a model would support the development, testing, and comparison of sensing algorithms without requiring rare measurement data.

To the best of our knowledge, only one such propagation model has been previously proposed, which assumes buried terrestrial fibres \cite{Virgillito:Oct23:earthquake_emulation_sensing,Awad:May24:environmental_surveillance_networks}.
This model obtains earthquake seismograms from a web service that synthesises seismograms (Syngine \cite{Krischer:Jul17:Syngine}), and integrates them into a numerical waveplate model that solves the Manakov-PMD (polarisation mode dispersion) equation \cite{Curti:Aug90:waveplate}.
However, the Manakov-PMD equation was derived in a reference frame that follows the rapid birefringence-induced SOP rotations at the carrier \cite{Marcuse:Sep97:Manakov_varying_birefringence}.
Therefore, strictly speaking the model in \cite[App.~A]{Awad:May24:environmental_surveillance_networks} predicts no SOP perturbations at the carrier frequency $\omega = 0$ during earthquakes.%
\footnote{%
    The authors of \cite{Curti:Aug90:waveplate} likely evaluated their model at an offset frequency $\omega \neq 0$.
}
Moreover, the model assumes a linear relation between ground motion and fibre strain, whereas submarine environments have a more complex transfer function \cite{Wang:Jun18:ground_motion_earthquake,Webb:Feb98:broadband_seismology_ocean}. 

In this paper, we present an end-to-end simulation model %
of the full interaction chain from earthquake-induced ground motion to optical signal propagation in submarine fibres. To the best of our knowledge, this is the first such model, and we demonstrate good qualitative alignment with measurements from \cite{Zhan:Feb21:polarization_wave_sensing}. 
In contrast to \cite{Awad:May24:environmental_surveillance_networks,Virgillito:Oct23:earthquake_emulation_sensing}, our model is based on the coupled Schr\"odinger equation (CSE) which resolves deterministic, time-dependent perturbations at the carrier.
Moreover, our implementation is open source%
\footnote{%
    \url{https://github.com/RM-8vt13r/quake-fibre}%
}, %
enabling reproducible sensing research and simulation of earthquake-induced fibre propagation without requiring experimental data.

\section{Earthquake-perturbed propagation model}
\begin{figure*}[t]
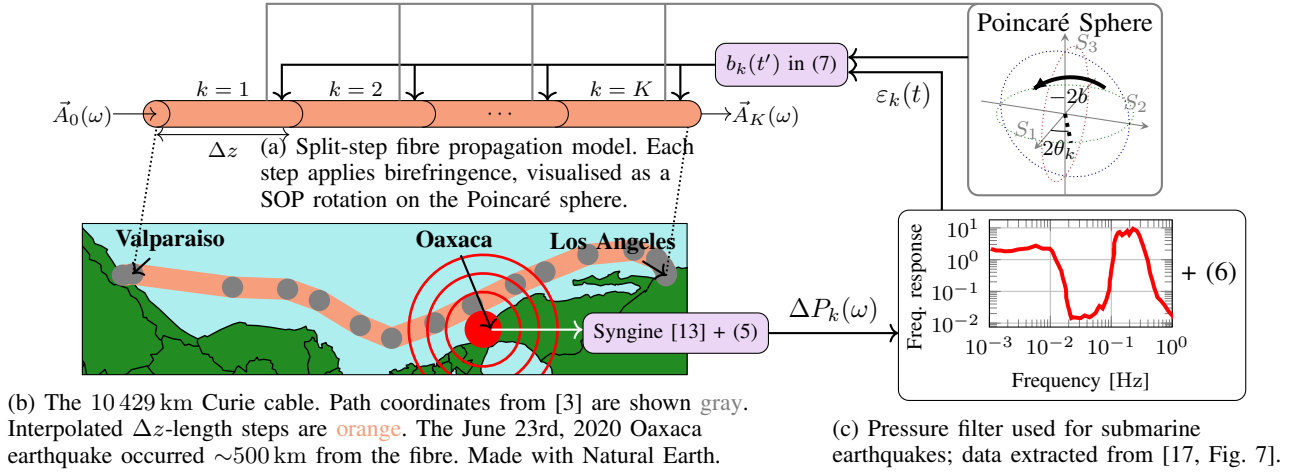

    \centering
    \newlength{\nodeDistance}
\newlength{\fibreWidth}
\newlength{\earthquakeWidth}
\newlength{\zoomVertexWidth}
\newlength\mapWidth
\newlength\fibreRadius
\newlength\sectionLength

\setlength{\nodeDistance}{.4cm}

\setlength\fibreRadius{.5em}
\setlength\sectionLength{5em}
\edef\sectionCount{4}
\pgfmathtruncatemacro{\sectionCountMinusOne}{\sectionCount - 1}
\pgfmathtruncatemacro{\sectionCountPlusOne}{\sectionCount + 1}

\setlength\mapWidth{8.1cm}
\edef\mapRotation{-125}

\setlength{\fibreWidth}{8pt}
\edef\fibreEdgeColour{Melon}
\edef\fibreVertexColour{gray}

\setlength\earthquakeWidth{28pt}
\edef\earthquakeColour{Red}
\edef\earthquakeRingCount{4} 
\edef\earthquakeRingWidth{1pt} 

\begin{tikzpicture}[
        x = 1cm,
        y = 1cm,
        node distance = \nodeDistance,
        field/.style = {
            align = center,
            anchor = center,
            inner sep = 0
        },
        zoom/.style = {
            field,
            draw = black,
            line width = .7pt
        },
        caption/.style = {
            field,
            anchor = north,
        },
        block/.style = {
            field,
            draw = black,
            minimum width = 4.5em,
            minimum height = 1.5em,
            rounded corners,
            font = \footnotesize
        },
        arrow/.style = {
            ->,
            draw = black
        },
        fibre/.style = {
            line width = \fibreWidth
        },
        earthquake/.style = {
            draw = \earthquakeColour,
            fill = \earthquakeColour,
            line width = \earthquakeRingWidth
        }
    ]
    
    \begin{scope}[local bounding box = Fibre, shift = {(0, 2.7)}]
        \input{figures/fibre}
    \end{scope}
    
    \begin{scope}[local bounding box = Map]
        \clip (0, -0.75) rectangle ++(8, 2);
        \begin{scope}[shift = {(2.5, 6.75)}, rotate = \mapRotation, transform shape]
            \input{figures/map}
        \end{scope}
    \end{scope}
    \draw (Map.north west) rectangle (Map.south east);
    
    \node[block, fill = DarkOrchid!20!white, inner sep = 4pt] (Syngine) at (7.85, -.2) {Syngine \cite{Krischer:Jul17:Syngine} + \eqref{eq:pressure}};

    \begin{scope}[local bounding box = Birefringence, shift = {(13, 2.7)}]
        \input{figures/birefringence}
    \end{scope}
    \draw[block, gray, thick] (Birefringence.north west) rectangle (Birefringence.south east);

    \begin{scope}[local bounding box = Filter, shift = {(12, -0.12)}]
        \input{figures/filter}
        \node[at = (FilterPlot.east), anchor = west] {+ \eqref{eq:strain}};
    \end{scope}
    \draw[block] (Filter.north west) rectangle (Filter.south east);

    \node[block, inner sep = 4pt, minimum width = 0, minimum height = 0, fill=DarkOrchid!20!white] (Strain) at (9.25, 3.35) {$b_k(t')$ in \eqref{eq:scaled_birefringence}};
    \draw[zoom, densely dotted] (Valparaiso) -- ($(Fibre1) - (0, \fibreRadius)$);
    
    \draw[zoom, densely dotted] (LosAngeles) -- ($(Fibre\sectionCountPlusOne) - (0, \fibreRadius)$);

    \draw[arrow, thick, white] (Oaxaca) -- (Oaxaca -| Syngine.west);
    
    \draw[arrow, thick] (Syngine) -- (Syngine -| Filter.west) node[midway, sloped, anchor = south] {$\Delta P_k(\omega)$}; 
    
    \foreach \index in {2,...,\sectionCountPlusOne}{
        \coordinate (StepCoordinate1) at ($(Fibre\index) + (-.3, \fibreRadius)$);
        \coordinate (StepCoordinate2) at ($(Fibre\index) + (-.1, \fibreRadius)$);
        \coordinate (BirefringenceCoordinate) at ($(Birefringence.north west)!.1!(Birefringence.north)$);
        \ifthenelse{\index = 2}{
            \draw[thick, gray] (StepCoordinate1) |- (BirefringenceCoordinate);
            \draw[thick, arrow] (Strain) -| (StepCoordinate2);
        }{
            \draw[thick, gray] (StepCoordinate1) -- (StepCoordinate1 |- BirefringenceCoordinate);
            \draw[thick, arrow] (Strain.west -| StepCoordinate2) -- (StepCoordinate2);
        }
    }

    \coordinate (StrainAnchorTop) at ($(Strain.north east)!.33!(Strain.south east)$);
    \coordinate (StrainAnchorBottom) at ($(Strain.north east)!.67!(Strain.south east)$);
    \draw[arrow, thick] (Birefringence.west |- StrainAnchorTop) -- (StrainAnchorTop);
    \draw[arrow, thick] ($(Filter.north west)!.25!(Filter.north)$) |- (StrainAnchorBottom) node[midway, anchor = north east] {$\varepsilon_k(t)$};

    \node[caption, at = (Fibre.south), shift = {(.8, .6)}, text width = .33\textwidth] {
        \subcaption{
            Split-step fibre propagation model.
            Each step applies birefringence, visualised as a SOP rotation on the Poincar\'e sphere.
        } \label{fig:system:fibre}
    };
    \node[caption, below = 0cm of Map, text width = .55\textwidth] {
        \subcaption{\raggedright
            The \qty{10429}{\kilo\meter} Curie cable.
            Path coordinates from \cite{submarine_cable_map} are shown \textcolor{\fibreVertexColour}{gray}.
            Interpolated $\Delta z$-length steps are \textcolor{\fibreEdgeColour}{orange}.
            The June 23rd, 2020 Oaxaca earthquake occurred $\sim$\qty{500}{\kilo\meter} from the fibre.
            Made with Natural Earth.
        } \label{fig:system:map}
    };
    \node[caption, at = (Filter.south), xshift = -.5em, text width = .33\textwidth] {
        \subcaption{
            Pressure filter used for submarine earthquakes; data  extracted from \cite[Fig.~7]{Webb:Feb98:broadband_seismology_ocean}.
        } \label{fig:system:filter}
    };
\end{tikzpicture}
    \caption{Block diagram of the end-to-end earthquake-perturbed propagation model.} \label{fig:system}
\end{figure*}
In the linear regime, the CSE describes how an optical carrier wave propagates through single-mode fibre according to \cite[Ch.~2]{Agrawal:2019:nonlinear_fiber_optics} and \cite{Marcuse:Sep97:Manakov_varying_birefringence}
\begin{equation} \label{eq:CNLSE}
    \jmath \frac{\partial\vec A(z, t')}{\partial z}
    + b \Sigma(\theta(z)){\vec A}(z, t') 
    = 0
\end{equation}
with $\jmath = \sqrt{-1}$, dual-polarisation complex signal envelope $\vec A(z, t') \in \mathbb{C}^2$, position $z$, time $t'$ relative to group velocity, half birefringence $b = \pi / L_\textrm{beat}$ at the carrier frequency, and beat leangth $L_\textrm{beat}$.
In \eqref{eq:CNLSE}, 
$
\Sigma(\theta(z)) = \left[
\begin{smallmatrix}
    \cos2\theta(z) & \sin2\theta(z) \\
    \sin2\theta(z) & -\cos2\theta(z)
\end{smallmatrix}
\right]
$
ensures that the birefringence is applied between the fastest- and slowest-travelling polarisations, where the fastest makes an angle $\theta(z)$ with a fixed reference frame.
In long-haul fibre systems, amplifiers add noise to the envelope during propagation.
We chose to omit attenuation and amplification, and model noise only at the end of the fibre, as we will describe in the next section.
As $\theta(z)$ is a stochastic process in $z$, \eqref{eq:CNLSE} has no closed-form solution.
The SOP is traditionally written as normalised real Stokes vector $\vec S = \begin{bmatrix}S_1 & S_2 & S_3\end{bmatrix}^T$ on the Poincar\'e sphere, as visualised in \cref{fig:system:fibre} (right).

To evaluate \eqref{eq:CNLSE} numerically, we follow \cite{Marcuse:Sep97:Manakov_varying_birefringence} and divide a fibre into $K$ $\Delta z$-length steps as visualised in \cref{fig:system:fibre} (left).
For each step $k \in \left\{1, \dots, K\right\}$, we generate $\theta_k = \theta\left(k\Delta z\right)$ as
\begin{equation}
    \theta_k - \theta_{k - 1} \sim \mathcal{N}\left(0, \frac{\Delta z}{2L_\textrm{corr}}\right),
\end{equation}
with correlation length $L_\textrm{corr}$ and $\theta_0 = 0$.
Similar to \cite[Eq.~23]{Marcuse:Sep97:Manakov_varying_birefringence}, we evaluate step $k$ as
\begin{align}
    &\vec{A}_k(\omega) =
    R_k
    D
    R_k^{-1}
    \vec{A}_{k - 1}(\omega), \label{eq:section_transfer_function} \\
    &R_k =
    \left[
    \begin{smallmatrix}
        \cos \theta_k & -\sin \theta_k \\
        \sin \theta_k & \cos \theta_k
    \end{smallmatrix}
    \right],
    D = e^{\circ \jmath b \Delta z
    \left[\begin{smallmatrix}
        1 & 0 \\
        0 & -1
    \end{smallmatrix}\right]
    }, \label{eq:section_transfer_function_matrices}
\end{align}
where $\vec A_k(t') = \vec A(k\Delta z, t')$, $\omega$ is frequency relative to the carrier and $e^\circ$ is the Hadamard exponential.

Now, we add the effect of earthquakes in \eqref{eq:section_transfer_function} as shown schematically in \cref{fig:system}.
First, we divide a fibre path \cite{submarine_cable_map} into steps of length $\Delta z$ using linear spline interpolation, as \cref{fig:system:map} shows.
Given measurements of a historic or hypothetical earthquake, Syngine \cite{Krischer:Jul17:Syngine} synthesises ground displacements in the centre of step $k$.
We retrieve historical earthquake measurements from the Global Centroid Moment Tensor database \cite{Ekstrom:Jun12:GCMT}.
Vertical ground displacement $u_{k}(t)$ at fibre step $k$, with absolute time $t$, creates water pressure differences \cite{An:Oct17:theoretical_solution_pressure}
\begin{equation} \label{eq:pressure}
    \Delta P_k(t) = \rho h \frac{\partial^2 u_{k}(t)}{\partial t^2},
\end{equation}
with water depth $h$, and water density $\rho$. 
We assume $h$ and $\rho$ constant.

The expression in \eqref{eq:pressure} was derived assuming that ocean water is incompressible.
In practice, this assumption breaks down above the $h$-dependent ocean resonance frequency \cite{Deng:Apr22:theoretical_pressure_seismic}.
\cite{Webb:Feb98:broadband_seismology_ocean} measured broadband seafloor pressure during seismic activity, demonstrating this frequency-dependence.
We assume that their pressure measurements represent the frequency response of the water column.
Therefore, we propose to model this frequency response by filtering $\Delta P_k(t)$ with the filter in \cref{fig:system:filter}, taken from \cite[Fig.~7b]{Webb:Feb98:broadband_seismology_ocean}.
Thus, our model relies on experimental measurements of the ocean frequency response. 

Finally, we obtain fibre strain \cite[Secs.~3 and 4]{Mecozzi:Jun21:polarization_sensing_submarine}
\begin{equation} \label{eq:strain}
    \varepsilon_k(t) = \alpha\xi \epsilon_P \Delta P_k(t),
\end{equation}
where $\Delta P_k(t)$ has been filtered as described, photoelasticity $\xi$ corrects for refractive index changes due to material compression, and $\epsilon_P$ scales pressure to strain.
Here, we introduced a calibration factor $\alpha$ to compensate uncertainties in $h$, $\rho$, $\epsilon_P$, and the pressure filter.
We assume that $\varepsilon_k(t)$ is stepwise constant in $t$ and replace $b$ in \eqref{eq:section_transfer_function_matrices} by \cite{Mecozzi:Jun21:polarization_sensing_submarine}
\begin{equation} \label{eq:scaled_birefringence}
    b_k(t') = \left(1 + \varepsilon_k(t' + \beta_1 k \Delta z)\right)b,
\end{equation}
where $t'$ is matched to $t$ using the polarisation-averaged group velocity $\beta_1$ \cite[Eq.~2.3.44]{Agrawal:2019:nonlinear_fiber_optics}.

\section{Simulation, results \& discussion}
\begin{table}[tb]
\centering
    \caption{
        Curie model parameters.
        We took $\xi$ and $\epsilon_P$ from \cite{Mecozzi:Jun21:polarization_sensing_submarine}, $L_\textrm{beat}$, $L_\textrm{corr}$ and $\Delta z$ from \cite{Marcuse:Sep97:Manakov_varying_birefringence}, and $\rho$ and $h$ from \cite[Fig.~3]{Xian:Jan23:density_seawater}.
        } \label{tab:fibre_parameters}
    \setlength{\tabcolsep}{4.5pt}
    \footnotesize
    \begin{tabular}{ll|ll}
    \toprule
    Symbol & Value & Symbol & Value \\
    \midrule
    $\beta_1$         & \qty{2.1e8}{\meter\per\second} & $K\Delta z$  & \qty{10429}{\kilo\meter} \\
    $L_\textrm{beat}$ & \qty{50}{\meter}               & $\rho$       & \qty{1045}{\kilo\gram\per\meter} \\
    $L_\textrm{corr}$ & \qty{100}{\meter}              & $h$          & \qty{4}{\kilo\meter} \\
    $\xi$             & \num{0.78}                     & $\epsilon_P$ & \qty{3.5}{\per\giga\pascal} \\
    $\Delta z$        & \qty{1.67}{\meter} & &  \\
    \bottomrule
    \end{tabular}
\end{table}

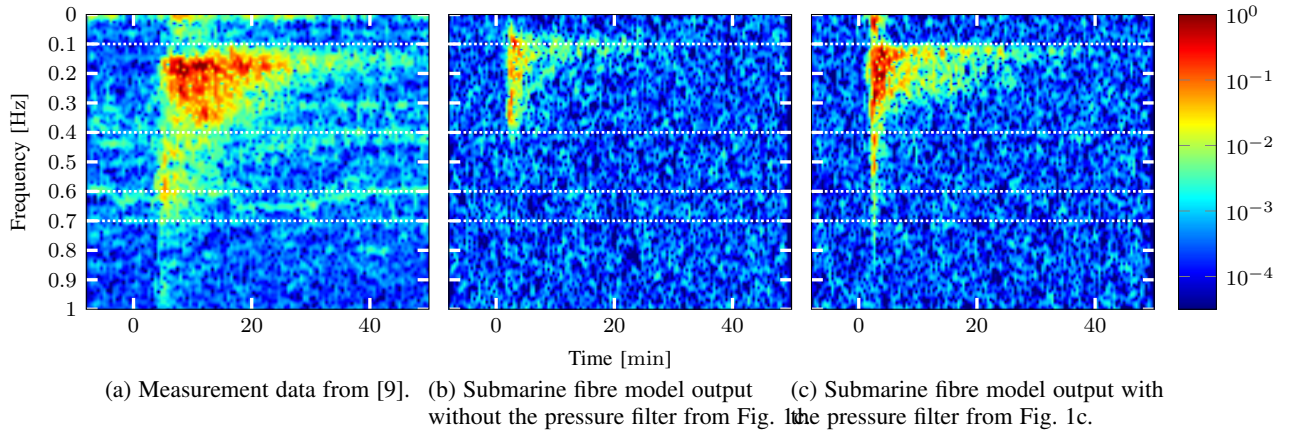
\begin{figure*}[tb]
    \centering
    \begin{tikzpicture}[
        font=\footnotesize,
        x = 1cm,
        y = 1cm,
        node distance = \nodeDistance,
        field/.style = {
            align = center,
            anchor = center,
            inner sep = 0
        },
        caption/.style = {
            field,
            anchor = north,
        }
    ]

    \pgfplotsset{
        set layers,
        spectrogram/.style = {
            scale only axis,
            font = \footnotesize,
            width = .25\linewidth,
            height = .215\linewidth,
            xmin = -8, xmax = 50,
            ymin = 0, ymax = 1,
            y dir = reverse,
            enlargelimits = false,
            axis on top,
            tick style = {
                white,
                line width = 1.25pt
            },
            ytick = {0, 0.1, ..., 1.1}
        },
    }

    \begin{axis}[
            spectrogram,
            name = Subplot1,
            ylabel = {Frequency [\unit{\hertz}]}
        ]
        
        \addplot graphics [
            xmin = -8.333, xmax = 66.667,
            ymin = 0, ymax = 10
        ] {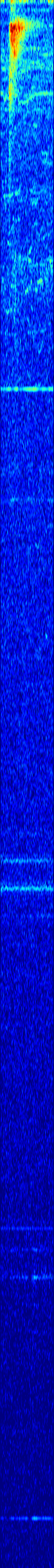};

        \coordinate (D) at (0, 0.1);
        \coordinate (F) at (0, 0.4);
        \coordinate (G) at (0, 0.6);
        \coordinate (H) at (0, 0.7);
    \end{axis}
    

    \node[caption, below = .7cm of Subplot1.south, text width = .28\linewidth] {
        \subcaption{
            Measurement data from \cite{Zhan:Feb21:polarization_wave_sensing}.
        } \label{fig:spectrograms:data}
    };

    \begin{axis}[
            spectrogram,
            name = Subplot2,
            at = {(Subplot1.right of north east)},
            anchor = {left of north west},
            xlabel = {Time [\unit{\minute}]},
            yticklabel = \empty
        ]
        
        \addplot graphics [
            xmin = -8.333, xmax = 50,
            ymin = 0, ymax = 10
        ] {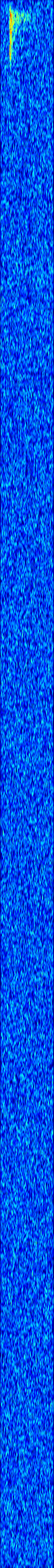}; 

    \end{axis}


    \node[caption, below = .7cm of Subplot2.south, text width = .28\linewidth, text depth = .75cm] {
        \subcaption{
            Submarine fibre model output without the pressure filter from \cref{fig:system:filter}.
        } \label{fig:spectrograms:unfiltered}
    };

    \begin{axis}[
            spectrogram,
            name = Subplot3,
            at = {(Subplot2.right of north east)},
            anchor = {left of north west},
            yticklabel = \empty,
            colorbar,
            colormap/jet,
            colorbar style = {
                name = bar,
                ytick = {-4, -3, -2, -1, 0},
                yticklabel = {$10^{\pgfmathprintnumber{\tick}}$},
            },
            point meta min = -4.5,
            point meta max = 0,
        ]
        \addplot graphics [
            xmin = -8.333, xmax = 50,
            ymin = 0, ymax = 10
        ] {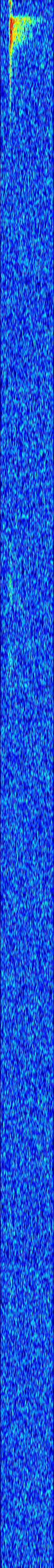}; 
        
    \end{axis}
    

    \node[caption, below = .7cm of Subplot3.south, text width = .28\linewidth] {
        \subcaption{
            Submarine fibre model output with the pressure filter from \cref{fig:system:filter}.
        } \label{fig:spectrograms:filtered}
    };

    \foreach \coord in {D, F, G, H}{
        \draw[white, line width = 1pt, densely dotted] (\coord -| Subplot1.west) -- (\coord -| Subplot3.east);
    }
\end{tikzpicture}
    \caption{
        Measured and modelled spectrograms, calculated over $S_1$ and $S_2$ and then summed. 
        The earthquake starts at \qty{0}{\minute}.
    } \label{fig:spectrograms}
\end{figure*}

We reproduce the experiment from \cite{Zhan:Feb21:polarization_wave_sensing} in simulation with the setup from \cref{fig:system,tab:fibre_parameters}, transmitting a dual-polarisation carrier wave at $\omega = 0$ (wavelength \qty{1550}{\nano\meter}).
We chose a small value of $\Delta z$ similar to \cite{Marcuse:Sep97:Manakov_varying_birefringence} to capture rapid birefringence variations, resulting in a simulation time of roughly \qty{6}{\hour} on an NVIDIA Tesla V100 GPU.
We set $\alpha = 1.52$ and add Gaussian noise to $\vec A(K\Delta z, t')$ to match the measured data as closely as possible, and then convert to Stokes parameters%
\footnote{%
    Note that \cite{Zhan:Feb21:polarization_wave_sensing} estimates the Stokes parameters from an equaliser in the receiver, instead of from $\vec A(K\Delta z, t')$ as we do.
}.

Like \cite{Zhan:Feb21:polarization_wave_sensing}, we align the SOP in Stokes space to the average $S_3$ over a \qty{200}{\second} moving average, which essentially acts as a high-pass filter.
Then, to visualise the effect of the earthquake on the fibre output, we follow \cite{Zhan:Feb21:polarization_wave_sensing} and sum the spectrograms of $S_1$ and $S_2$.
We generate spectrograms using a \qty{102.4}{\second} Blackman-Harris window, \qty{75}{\percent} window overlap and \qty{409.6}{\second} zeropadded Fourier transforms.
The result is shown in \cref{fig:spectrograms}, where \cref{fig:spectrograms:data} shows the result from \cite{Zhan:Feb21:polarization_wave_sensing}, and \cref{fig:spectrograms:unfiltered} and \cref{fig:spectrograms:filtered} show our model output excluding- and including the pressure filter from \cref{fig:system:filter}.
The spectrograms visualise the amplitude of the fibre output SOP rotations, as a function of time and frequency.
The results shown are for one random realisation of $\theta_k$.
We also verified that different realizations produce very similar spectrograms, with only a minor amplitude change near \qty{0}{\minute}.

Both in the measurements and our filtered model, the SOP fluctuates in the \qtyrange{0}{0.7}{\hertz} band when the earthquake begins near \qty{0}{\minute}.
For the unfiltered model, this is \qtyrange{0.05}{0.4}{\hertz}.
In both the measurement and our filtered model, the SOP rotations are strongest between \qtylist{0.1; 0.4}{\hertz}.
In this range, the model output shows larger spectrogram amplitudes than the experimental measurements.
This amplitude is prone to our choice of $\alpha$, and the strong pressure amplification by the filter from \cref{fig:system:filter} in this band.

After the earthquake reaches the fibre, the SOP fluctuations diminish over time in all spectrograms.
Here, high-frequency components attenuate quickly whereas low frequencies persist for up to \qty{40}{\minute}.
We observe that the measured fluctuations attenuate slowly, with the strongest frequency component reaching an amplitude of $10^{-2}$ after roughly \qty{30}{\minute}.
In the unfiltered and filtered models, this point is reached after \qtylist{5; 15}{\minute} already.
In addition, we observe a bandgap in the measurements and filtered model from \qtyrange{0.02}{0.1}{\hertz}.
The unfiltered model shows a gap up to \qty{0.03}{\hertz}.

The measured spectrogram shows harmonic components around \qtylist{0; 0.4; 0.6}{\hertz}, which are absent in both modelled spectrograms.
As these harmonics are present before and after the earthquake, we conjecture that they are caused by another external event, such as ocean waves.



Overall, the modelled and measured perturbed SOPs are qualitatively aligned.%
\footnote{%
    Normalised cross-correlation between the spectral centroids \cite{Mousavi:Oct16:seismic_features_discrimination} of our filtered model and the measurements yields \num{0.70}, showing a good quantitative match as well.
}
Both the shape of the spectrogram patterns, and the low-frequency bandgap, match.
There is a mismatch in amplitude, which is modelled larger than measured at the start of the earthquake, and attenuates faster. 
This mismatch may be partially explained by the uncertainty in several model parameters. 
For example, the true values for $h$ and $\rho$ along the fiber, as well as the birefringence properties of the Curie cable are not known exactly. 
Moreover, our model assumes that changing water pressure is the primary SOP modulator as in \cite{Mecozzi:Jun21:polarization_sensing_submarine}.
However, \cite{Mecozzi:Jun21:polarization_sensing_submarine} also notes that ``for earthquake excitations [...], other mechanisms for SOP modulation cannot be excluded.''
Besides uncertainty in the model parameters, the mismatch may therefore also reflect the presence of an additional coupling mechanism.
In particular, as submarine cables partially lie on or in the seafloor, resulting longitudinal cable pulling may also add to the observed SOP modulation.

\section{Conclusion}
We presented a novel end-to-end model of earthquake-perturbed submarine fibre signal propagation.
We verified the model on experimental measurements from the Curie cable during the 2020 Oaxaca earthquake. 
We believe that our model can aid large-scale design and testing of earthquake fibre sensing algorithms 
and help planning the placement of new sensing fibres.

The work presented here is a first step in the direction of a more complete open-source end-to-end modelling of earthquake sensing with optical fibres.
Future work includes 
the inclusion of an analytical pressure filter \cite{Deng:Apr22:theoretical_pressure_seismic}, and the addition of SOP modulation by longitudinal fibre pulling.

\clearpage
\section{Acknowledgements}
This work was partially supported by the Swedish Research Council under grants no.~2025-04836 and 2025-03824, by the project PINTO with file number NGF.1609.242.015 of the National Growth Fund (NGF) AiNed programme which is financed by the Dutch Research Council (NWO), and by the European Union’s Horizon 2024 research and innovation program under the Marie Skłodowska-Curie grant agreement No 101119983.

\bibliographystyle{IEEEtran}
\bibliography{references}

\end{document}